\documentclass[twocolumn,tighten]{aastex631}

\usepackage{xcolor}
\definecolor{mygreen}{rgb}{0.19,0.55,0.11}

\shorttitle{Accretion induced collapse of neutron stars}
\shortauthors{Chen et al.}
\graphicspath{{./}{figures/}}

\begin{document}

\title{Does Nature allow formation of ultra-compact black hole X-ray binaries\\via accretion-induced collapse of neutron stars?}

\correspondingauthor{Hai-Liang Chen}
\email{chenhl@ynao.ac.cn}

\author{Hai-Liang Chen}
\affiliation{Yunnan Observatories, Chinese Academy of Sciences (CAS), Kunming 650216, P.R. China}
\affiliation{Key Laboratory for the Structure and Evolution of Celestial Objects, Chinese Academy of Sciences, Kunming 650011, China}
\affiliation{International Centre of Supernovae, Yunnan Key Laboratory, Kunming 650216, P. R. China}

\author[0000-0002-3865-7265]{Thomas M. Tauris}
\affiliation{Department of Materials and Production, Aalborg University, Skjernvej 4A, DK-9220~Aalborg {\O}st, Denmark}

\author{Xuefei Chen}
\affiliation{Yunnan Observatories, Chinese Academy of Sciences (CAS), Kunming 650216, P.R. China}
\affiliation{Key Laboratory for the Structure and Evolution of Celestial Objects, Chinese Academy of Sciences, Kunming 650011, China}
\affiliation{University of the Chinese Academy of Sciences, Yuquan Road 19, Shijingshan Block, 100049, Beijing, China}
\affiliation{International Centre of Supernovae, Yunnan Key Laboratory, Kunming 650216, P. R. China}

\author[0000-0001-9204-7778]{Zhanwen Han}
\affiliation{Yunnan Observatories, Chinese Academy of Sciences (CAS), Kunming 650216, P.R. China}
\affiliation{Key Laboratory for the Structure and Evolution of Celestial Objects, Chinese Academy of Sciences, Kunming 650011, China}
\affiliation{University of the Chinese Academy of Sciences, Yuquan Road 19, Shijingshan Block, 100049, Beijing, China}
\affiliation{International Centre of Supernovae, Yunnan Key Laboratory, Kunming 650216, P. R. China}
\begin{abstract}
The formation path to ultra-compact X-ray binaries (UCXBs) with black hole (BH) accretors is still unclear. In the classical formation scenario, it is difficult to eject the massive envelope of the progenitor star of the BH via common envelope process. Given that some neutron stars (NSs) in binary systems evidently have birth masses close to $\sim 2.0\;M_\odot$, we explore here the possibility that BH-UCXBs may form via accretion-induced collapse (AIC) of accreting NSs, assuming that these previously evolved in LMXBs to masses all the way up to the maximum limit of a NS. We demonstrate this formation path by modelling a few cases of NS-UCXBs with initial NS masses close to the maximum mass of a NS that evolve into BH-UCXBs after the NS accretes material from its He~WD companion. We follow the evolution of the post-AIC BH-UCXB and, based on simple arguments, we anticipate that there is about one BH-UCXB with an AIC origin and a He~WD donor within the current sample of known UCXBs and that 2--5 such BH-UCXBs may be detected in gravitational waves by LISA. In addition, we find that the X-ray luminosity of NS-UCXBs near their orbital period minimum exceeds $\sim 10^{39}\;{\rm erg\;s^{-1}}$ and thus such systems may appear as ultraluminous X-ray sources.
\end{abstract}


\section{Introduction} \label{sec:intro}

Ultra-compact X-ray binaries (UCXBs) are X-ray binaries with short orbital periods ($P_{\rm orb} < 80\;$min). The accretors are neutron stars (NSs) or black holes (BHs) and the companion stars are hydrogen-deficient, partially or fully degenerate low-mass ($\la 0.2\;M_\odot$) stars \citep{rjw82,sdv86}. 
UCXBs are important for studies of binary evolution \citep[e.g.][]{py14,cthc21,tv23}, accretion disks \citep[e.g.][]{hiep+13}, and accreting millisecond pulsars \citep[e.g.][]{wijn10}. Due to their short orbital periods, they are also very important sources for gravitational wave (GW) detectors, such as LISA \citep{aaaa+23}, TianQin \citep{lcdg+16}, and Taiji \citep{rgcz20}.

So far, around 20 confirmed and candidate UCXBs have been found in our Galaxy \citep{ijm07,nj10}. The far majority of accretors in this sample of systems are NSs. Only one UCXB is confirmed to have a BH accretor \citep[the source X9 in the globular cluster 47~Tucanae,][]{bhtm+17}. In addition, there is another BH-UCXB candidate found in the globular cluster RZ2109 in the elliptical galaxy NGC~4472 \citep{szmk+14}. 
These two BH-UCXBs are remarkable as they show signs of oxygen enrichment, pointing to a possible CO white dwarf (WD) donor star. 

The formation of BH-UCXBs remains an unsolved question. In globular clusters, BH-UCXBs may simply be produced by dynamical interactions that pair the BH with its companion star \citep[e.g.][]{irld+05,icfh+10}. In the Galactic field, however, the evolution of binary systems is challenging to model. One reason is the removal of the massive envelope of the BH progenitor star by its companion star, either via a common envelope (CE) phase or stable mass transfer via Roche-lobe overflow (RLO), while at the same time producing a tight-orbit system. Depending on the ZAMS masses of the two stellar components in the original binary system, the former (CE) scenario may prove difficult because of the large binding energy of the relatively massive envelope \citep{pve97,kalo99,prh03,ijcd+13,ktl+16}. The latter (RLO) scenario has been briefly investigated in \citet{tmk+18} using semi-numerical modelling. 

In globular clusters, however, a subclass of tight NS+WD systems with very small eccentricities cannot be the direct product of an exchange interaction. Currently, two dozen of radio millisecond pulsars are found in globular clusters with a WD companion and an eccentricity, $e<0.001$ \citep[cf. ATNF Pulsar Catalogue][]{mhth05}.
This indicates that the progenitors of these NS+WDs either evolved in complete isolation, or a NS was captured in a binary via dynamical interactions {\em before} the system underwent RLO, producing the WD in a circular orbit. 
The exact origin of the tight NS+WD system is nevertheless irrelevant for our computations, which follow the evolution of the subsequent UCXB phase, once the WD is forced to fill its Roche lobe because of gravitational damping of the orbit.

In general, UCXBs with a NS/BH accretor have been suggested to be a product of binary star evolution in one of the following three scenarios (depending on the nature of the companion star): i) a WD companion that loses orbital angular momentum because of GW radiation, leading to the onset of mass transfer as mentioned above \citep{pw75,ynv02,stli17,qjc23}; ii) a helium star donor filling its Roche lobe \citep{sdv86,yung08,wcl+21}; and finally, iii) a binary consisting of a low-mass main-sequence (MS) star that initiates mass transfer when the donor star is near the end of hydrogen core burning, thereby being stripped to a naked He-rich core \citep{nrj86,prp02}.
If companions to BH-UCXBs turn out to be He~WDs, or naked He-rich cores evolved from stripped low-mass MS stars, their formation path is likely to follow that of NS-UCXBs with similar donors --- a scenario that despite common acceptance still challenges our current understanding of how magnetic braking operates to keep the orbit tight by removal of orbital angular momentum \citep{cthc21}.

\subsection{Aim, motivation and rationale of this study}
\label{subsec:aim}

Here we investigate an alternative formation scenario for BH-UCXBs, in which an accreting NS in an UCXB reaches a critical mass ($M_{\rm NS,max}$)\footnote{ In this paper, when talking about the masses of NSs/BHs, we refer to their gravitational masses.} and undergoes accretion-induced collapse (AIC) to produce a BH --- in a process that is analogous to a massive (ONeMg) WD accumulating material until reaching the Chandrasekhar mass limit ($\sim 1.4\;M_\odot$) and imploding to leave a NS remnant. The AIC formation model for NSs was suggested already long ago by theorists \citep[for early works, see e.g.][]{cil90,nsky+91,nk91,tw92}. Specifically for low-mass X-ray binaries (LMXBs), the AIC NS formation model was suggested already in the mid-1970s by \citet{vdh76,vdh77}, and further developed by other groups. For a review on AIC and its astrophysical evidence, see \citet{tv23}.

A second motivation for investigating NS-AIC is the apparent evidence of low-mass BHs (i.e. the so-called lower-mass gap BHs with masses $2-5\;M_\odot$), discovered by the LIGO-Virgo-KAGRA network of GW detectors \citep{aaa+21c} and also alluded to earlier in a study of the mass distribution of Galactic X-ray binaries \citep{bjco98}. 
Of course, if such BHs were formed via NS-AIC then a GW merger event of two BHs would require either a triple-star origin or subsequent dynamical pairing of two BHs (one being the low-mass BH formed via NS-AIC) in a dense stellar environment \citep{rcr16}. 

Our rationale for investigating NS-AIC as a formation path to low-mass BHs is driven by accumulating empirical evidence for the existence of very massive NSs in binary systems, close to the putative maximum mass limit of a NS, $M_{\rm NS,max}$. 

Regarding the maximum mass of a NS, it is still under debate and strongly depends on the equation of state (EOS) \citep[e.g.][]{lp07,sfh+2017,grb21}.
 On the theoretical side, $M_{\rm NS,max}$ has traditionally been considered up to $3\;M_{\odot}$ \citep[e.g.][]{rr74,kb96}. However, statistics of the radio pulsar and/or the X-ray NS mass distribution tend to converge to an upper limit close to $2.20\;M_\odot$ \citep{ato+16,asb18,fc20}. Especially after the first GW detection of a NS+NS merger event \citep[GW170817,][]{aaa+17} new constraints have been derived of $M_{\rm NS,max}=2.16^{+0.17}_{-0.15}\;M_\odot$ \citep[][]{rmw18}; see also \citet{mm17,mpwr20}.
 However, according to some studies, a larger value of $M_{\rm NS,max}$ is still possible. For further discussions, and even the possibility of $M_{\rm NS,max}\sim 2.6\;M_\odot$, see e.g. \citet{grb21} and references therein.

Empirical evidence for the highest NS masses of binary millisecond pulsars (MSPs) are PSRs~J1614-2230 \citep{dprr+10,fpes+16,abbc+18}, J0348+0432 \citep{afwt+13}, J0740+6620 \citep{cfrd+20} and J2215+5135 \citep{lsc18}, with inferred masses of: $1.908\pm0.016\;M_{\odot}$, $2.01\pm0.04\;M_{\odot}$, $2.14^{+0.10}_{-0.09}\;M_{\odot}$ and $2.27\pm0.17\;M_{\odot}$, respectively.
Furthermore, \citet{lina20} presented evidence that MSPs in some redback and black widow systems may have NS masses larger than $2.0\;M_{\odot}$, although the inferred NS masses in some of these systems may suffer from systematic effects; and most recently \citet{rkfb+22} claimed a NS mass in PSR~J0952-0607 of $2.35 \pm 0.17\;M_{\odot}$. 
As a result of inefficient LMXB accretion \citep[e.g.][]{ts99,avkf+12}, little mass is accumulated  by the NSs during the evolution of LMXBs in which NSs are recycled to MSPs. Therefore, these MSPs listed above are likely to have original NS birth masses relatively close to their current masses \footnote{It has been demonstrated that a NS needs only accreting a few $0.01\;M_\odot$ to produce an MSP \citep{tlk12}. 
Note also that some young NSs in HMXBs, e.g. Vela~X-1, IGR~J17252$-$3616 and 4U~1700$-$377 \citep{mwb12,fbl+15}, evidently had a birth mass near $1.8-2.1\;M_\odot$.}. 

In any case, the important fact here is that MSPs are found with very massive NSs and that these masses would be similar to pre-UCXB NS masses.
It is therefore tantalizing now to investigate the possibility of NS-AIC events, and the most promising systems in which such events can occur are NS-UCXBs. 
For a recent independent investigation of low-mass BHs produced via AIC in LMXBs,  see \citet{gls22} and \citet{skkb+22}; and see Section~\ref{subsec:com} for further comparison to their results.

For this study, we shall assume that some NSs in pre-UCXBs are produced with masses close to the $M_{\rm NS,max}$ limit, irrespective of its exact value. For our computations, we simply applied $M_{\rm NS,max}=2.20\;M_\odot$, although the value is uncertain and may possibly be larger, e.g. $M_{\rm NS,max}=2.60\;M_\odot$. 
Independent of this, one aspect of this study is that we investigate how close to $M_{\rm NS,max}$ the initial pre-UCXB NS masses have to be in order to produce BH-UCXBs via AIC.

The AIC of a NS to a BH may produce many interesting electromagnetic signatures. It has been suggested that NS-AICs is a possible mechanism for both long and short gamma-ray bursts \citep[e.g.][]{vs99,mrz05,da06}; and they have also been suggested to be progenitors of non-repeating fast radio bursts \citep{fr14}.

In our previous work \citep{ctch22b}, we have modeled the evolution of NS-UCXBs consisting of an accreting NS and a He~WD donor star. During the evolution with mass transfer (initially at a high rate), the NS may reach its limiting maximum mass and implode to form a BH at some point. The precise aim of the work presented here is a first study on the possibility and the limitations of the formation of BH-UCXBs via AIC of a NS in UCXBs with a He~WD donor and investigate their post AIC observational properties. 

The rest of the paper is organized as follows. In Section~\ref{sec:met}, we briefly introduce the method and assumptions adopted in our simulations. In Section~\ref{sec:res}, we present the results obtained from our simulations. Finally, we discuss our results in Section~\ref{sec:discussion} and summarize our conclusions in Section~\ref{sec:conclusion}.

\section{Modelling of binary evolution}\label{sec:met}

\subsection{Binary stellar evolution calculations}\label{subsec:bec}
In our work, we adopt the stellar evolution code Modules for Experiments in Stellar Astrophysics (\textsc{mesa}, version 12115, \citealt{pbdh+11,pcab+13,pmsb+15,psbb+18,pssg+19}) to model the evolution of binaries\footnote{ The \textsc{mesa} inlists and output files used in our study are available on the \dataset[Zenodo community]{https://zenodo.org/communities/mesa/}.}. The UCXB simulations include three phases: 1) pre-AIC phase --- evolution of NS+He~WD binary; 2) AIC implosion --- NS collapse into a BH, leading to instantaneous widening of the orbit; 3) post-AIC phase --- evolution of BH+He~WD binary. 
Below we briefly summarize the main assumptions adopted in our simulations. 

During the UCXB evolution in the pre-AIC phase, the amount of material accreted by the NS is rather modest because the mass is transferred from the WD donor at a rate exceeding the Eddington limit of the accreting NS ($\sim 3\times 10^{-8}\;M_\odot\,{\rm yr}^{-1}$) --- initially by more than an order of magnitude (Figs.~\ref{fig:aic_ex1} and \ref{fig:aic_ex2}, top panels). Therefore, to trigger an AIC event and formation of a BH, the initial NS has to be massive (close to $M_{\rm NS,max}$) at the beginning of the pre-AIC phase. Thus, without considering the formation processes of NS+He~WD binaries, we simply concentrate on AIC candidates and assume at this evolutionary stage NS masses within $0.01-0.04\;M_\odot$ of $M_{\rm NS,max}$, because we previously found that NSs in UCXBs with He~WD donors accrete at most $\sim 0.037\;M_\odot$ \citep{ctch22b}; see also Section~\ref{subsec:number_BH-UCXBs} for a confirmation. 
That is, following the discussion on the maximum mass of a NS in Section~\ref{subsec:aim}, we conservatively assume the maximum NS mass to be $M_{\rm NS,max}=2.20\;M_{\odot}$ and thus start our default computations here with a NS of $2.18\;M_{\odot}$ to ensure AIC.

For the initial WD masses, we choose two representative cases of 0.17 and $0.40\;M_{\odot}$, respectively. The former case represents NS+He~WD UCXBs produced via stable RLO in LMXBs. For such systems it has been demonstrated that, due to the orbital period--WD mass correlation \citep[e.g.][]{ts99}, the He~WDs all have masses of $0.15-0.17\;M_\odot$ \citep{taur18,cthc21}. The latter case with a $0.40\;M_\odot$ WD donor represents an example of an UCXB produced from a binary system that underwent CE evolution. 
For the structure of the WDs, we adopt the models from \citet{ctch22b}, in which WD models are produced from the evolution of LMXBs. The initial central temperature of the He~WDs applied here are close to $10^{7}\;$K. The initial orbital periods were set to 0.30 and $0.02\;$days, respectively. (We checked that the choice of initial orbital periods does not influence our results significantly.) 

We make use of the \textsf{star\_plus\_point\_mass} test suite in \textsc{mesa} to model the evolution of UCXBs. The (pre-AIC) NS and (post-AIC) BH are treated as point masses. The orbital angular momentum losses due to both GW radiation and mass loss from the system are taken into consideration. 
The orbital angular momentum loss due to GW radiation is computed as \citep{ll71}:
\begin{equation}
	\frac{{\rm d}J_{\rm gw}}{{\rm d}t} = -\frac{32}{5}\frac{G^{7/2}}{c^{5}}\frac{M^2_{\rm a}M^2_{\rm WD} (M_{\rm a}+M_{\rm WD})^{1/2}}{a^{7/2}}\;,
\end{equation}
where $M_{\rm a}$ is the mass of the accretor, i.e., NS or BH; $M_{\rm WD}$ is the mass of the WD donor; $a$ is the binary separation; $G$ and $c$ are the gravitational constant and the speed of light in vacuum, respectively. 

In our calculations, we adopt the \textsf{Ritter} scheme to compute the mass-transfer rate \citep{ritt88} and the isotropic re-emission model to compute the orbital angular momentum loss due to mass loss \citep{tv06,tv23}. In the isotropic re-emission model, the material is assumed to be conservatively transferred from the WD to the accretor, from where a fraction ($\beta$) of material is lost from the system, taking away the specific orbital angular momentum equal to that of the accretor. We assume here $\beta = 0.70$ \citep{ts99,avkf+12} for the pre-AIC phase and $\beta = 0.0$ for post-AIC phase. 

In both the pre-AIC and the post-AIC stages, the accretion rate onto the NS and BH, respectively, is assumed to be restricted by the Eddington limit: 
\begin{equation}
    \dot{M}_{\rm Edd} = \frac{4\pi\,GM_{\rm NS}}{\eta \,0.20\,(1+X) c}\;,
\end{equation}
where $X$ is the hydrogen mass fraction of the accreted material. In our He~WD donor star models, the He~WDs have thin H shells; and thus shortly after the onset of the UCXB phase, $X = 0$. In our calculations, we assume the efficiency of the NS of converting accreted rest mass to radiative energy to be $\eta = 0.15$ \citep{lp07}. Following \citet{prh03}, the efficiency of the BH in converting accreted rest mass to radiative energy is approximately given by: 
\begin{equation}
\eta=1-\sqrt{1-\left(\frac{M_{\mathrm{BH}}}{3 M_{\mathrm{BH}}^{0}}\right)^{2}},
\end{equation}
for $M_{\rm BH} < \sqrt{6}\;M^{0}_{\rm BH}$. Here, $M^{0}_{\rm BH}$ and $M_{\rm BH}$ are the initial and present masses of the BH, respectively. In our calculations, it is always the case that $M_{\rm BH} < \sqrt{6}\;M^{0}_{\rm BH}$.  More specifically, the accretion rate of the NS/BH is given by \citep[see e.g. eq.(13) in][]{tsyl13}:
\begin{equation}
  \dot{M}_a = (1-\beta)(1-\eta)\left( |\dot{M}_{\rm WD}| -\max \left[|\dot{M}_{\rm WD}|-\dot{M}_{\rm Edd}\;,0\right] \right)
\end{equation}

Following the discussions in Sections~\ref{subsec:aim} and \ref{subsec:bec}, we assume $M_{\rm NS,max}=2.20\;M_{\odot}$.
If the computed mass of the accreting NS in our evolutionary sequences exceeds this critical value, we assume that the NS will collapse into a BH with a mass of $2.00\;M_{\odot}$ (i.e. during the AIC implosion, we assume that an equivalent mass of $0.20\;M_{\odot}$ is lost as released gravitational binding energy via neutrinos). 

There is a consensus in the community that the physics of an AIC event is not very different from that of an electron-capture supernova, and it is therefore expected that also NSs formed via AIC will receive a small kick \citep[if any significant kick at all,][]{kjh06,dbo+06}. As a result, NSs formed via AIC will have a high chance of remaining inside a globular cluster where the escape velocity is $<50\;{\rm km\,s}^{-1}$.
In analogy, we assume in our simulations that the natal kick of the BH born in an AIC event of a NS is between 0 and $50\;{\rm km\,s}^{-1}$. 

\subsection{Kinematic effects of the AIC event}
Regarding the kinematic effect of the AIC event, we follow the prescription of \citet{tsyl13} (see their section~3.2.1) to study the effects of sudden mass loss and natal kicks of BHs in post-AIC binary systems. From this prescription, we calculate the binary separation and eccentricity just after the AIC event, assuming here for simplicity $\theta = \phi = 0$ for the direction of the BH kick. 
Of course, a large parameter space of post-AIC orbits are possible depending on the value of a possible natal kick, $\vec{w}$, i.e. its magnitude and direction: ($w$, $\theta$, $\phi$). However, the post-AIC evolution is to a large extent mainly driven by loss of orbital angular momentum due to GWs. Thus wider post-AIC orbits will simply just require more time to decay back and follow the same evolution as expected for more tight post-AIC orbits. Furthermore, AIC kicks are expected to be small (see previous paragraph) and also each post-AIC evolution calculation requires a significant amount of computational time. Because the aim of this work is to validate AIC of a NS as a possible (but rare) formation path to produce low-mass BHs, we therefore proceed with just a few kick (or no kick) cases. Our main conclusions remain invariant to the exact AIC kick.

We check if the post-AIC binary system will experience mass transfer at periastron passage. If this is the case, we assume that the binary will merge and disregard any further calculation of that system. Otherwise, we assume that the binary system circularizes with time due to tidal interactions after the AIC event. To be specific, we assume that the post-AIC semi-major axis will be reduced by a factor of $(1 - e^{2})$ ($e$ is the eccentricity just after AIC) from conservation of orbital angular momentum \citep{sut74}. With this prescription, we can continue the simulation of the post-AIC stage and calculate through the subsequent UCXB phase with the BH accretor until the He~WD remnant is stripped down to a value of $\sim 0.01\;M_\odot$ after several Gyr. The results from these simulations are presented in the following section.

\section{Results}
\label{sec:res}

\begin{figure*}
    \centering
    \includegraphics[width=\columnwidth]{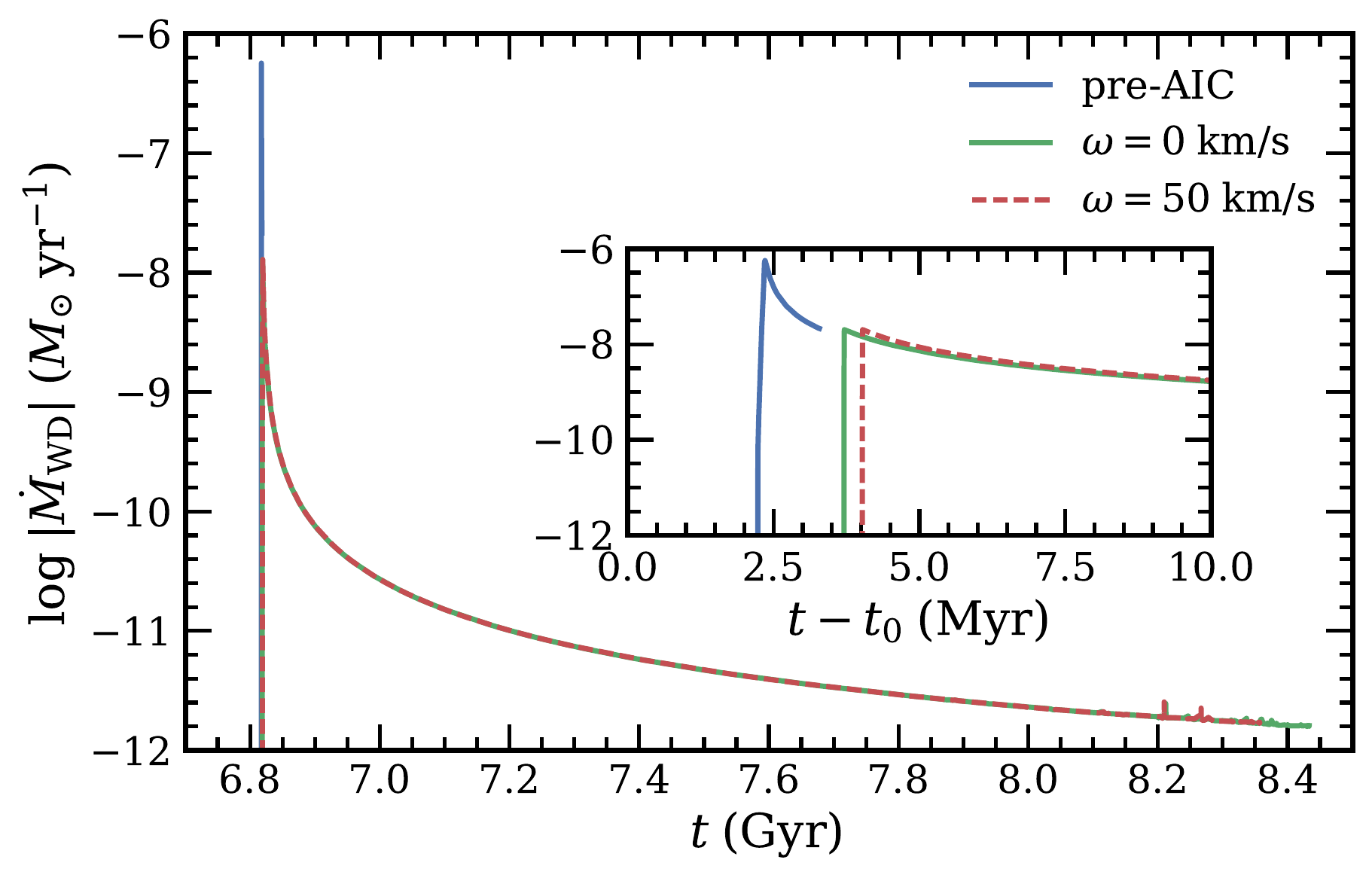}
    \includegraphics[width=\columnwidth]{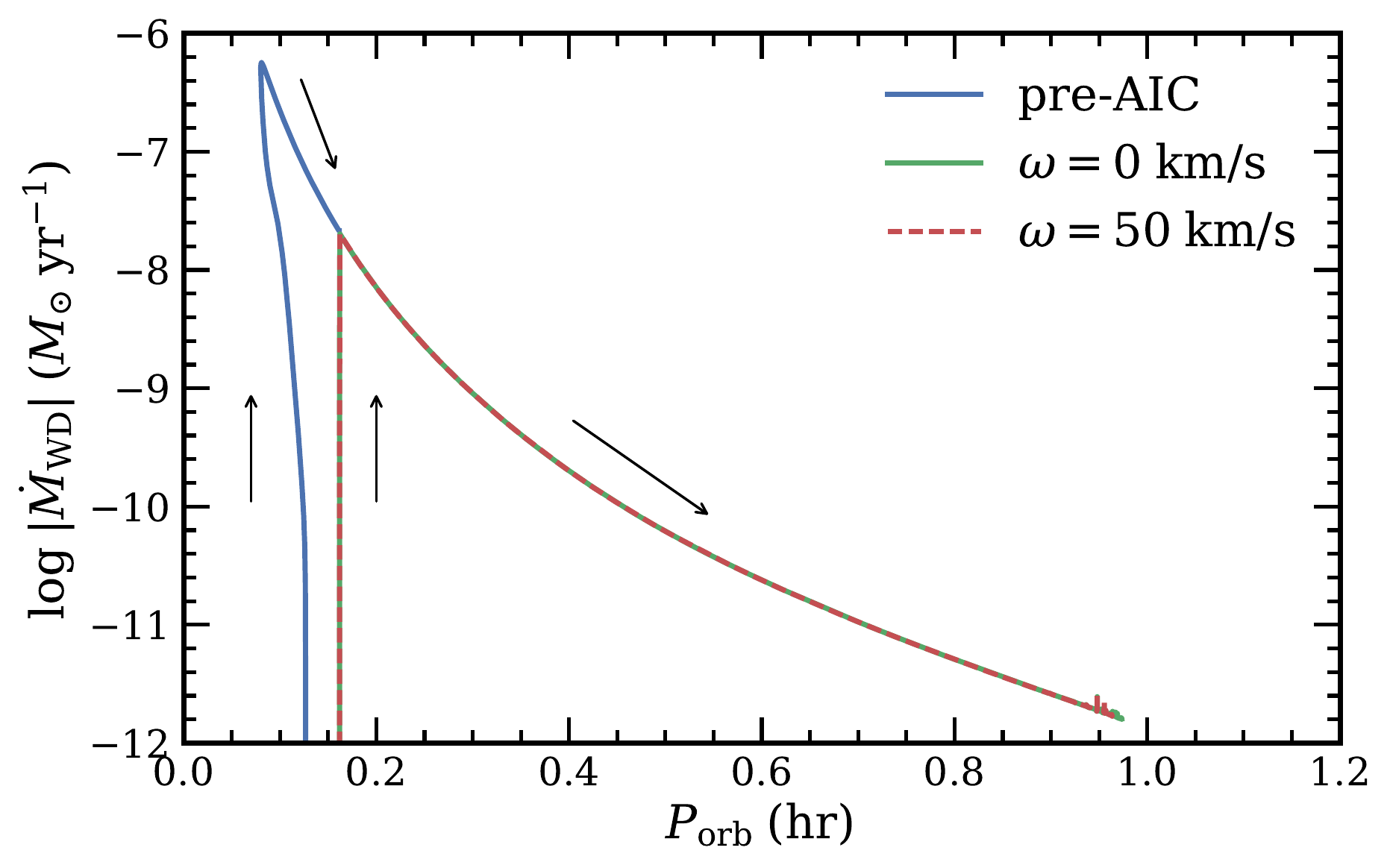}
    \includegraphics[width=\columnwidth]{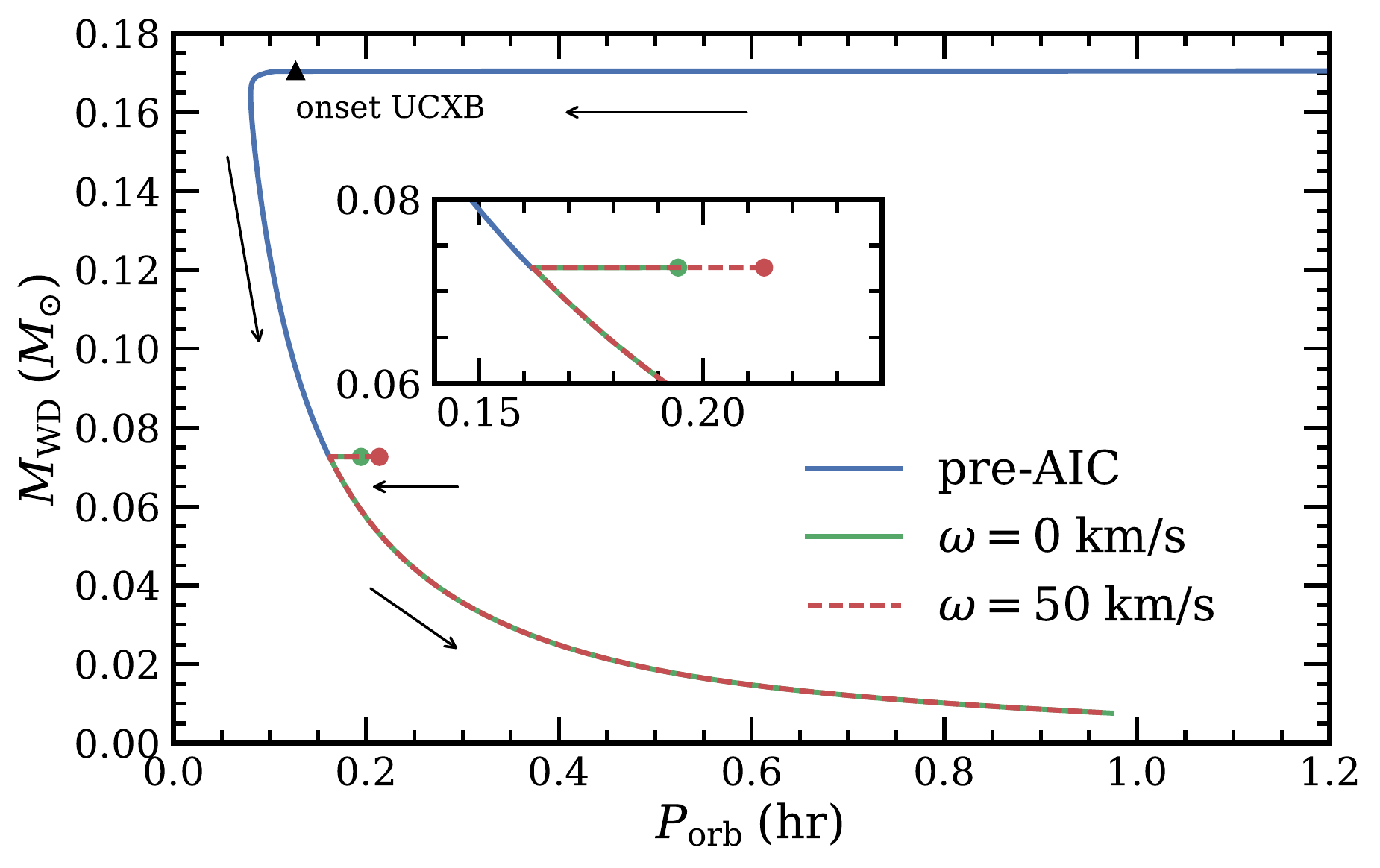}
    \includegraphics[width=\columnwidth]{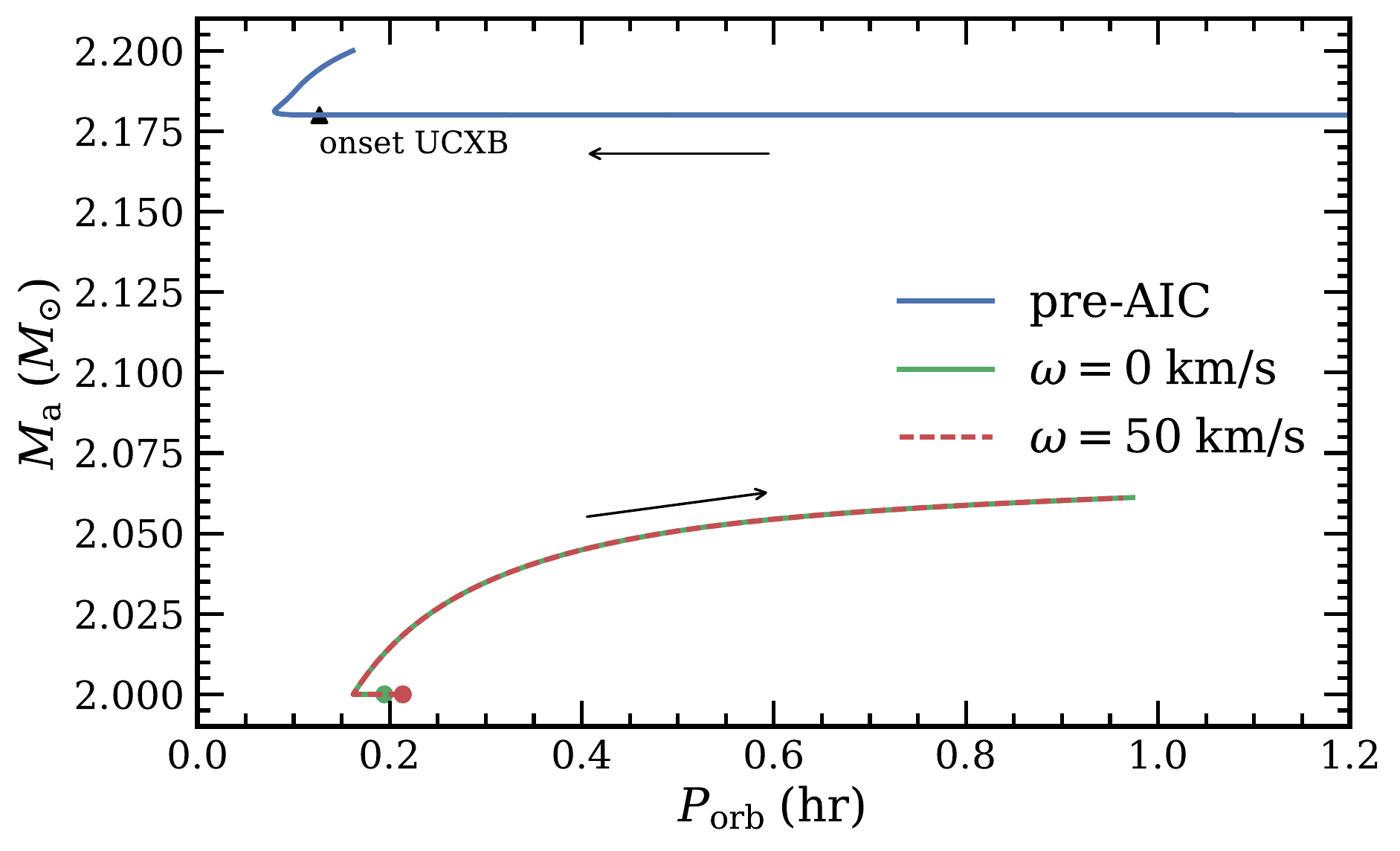}
    \caption{Evolution of mass-transfer rate as a function of time (upper-left panel) including an inset showing the early mass-transfer phase with a time offset of $t_{0} = 6.815\;{\rm Gyr}$, marking the onset of the UCXB phase. Evolution of mass-transfer rate (upper-right panel), WD mass (lower-left panel) and accretor mass (lower-right panel) as a function of orbital period. The initial binary parameters in this example are $M_{\rm a} = 2.18\;M_{\odot}$, $M_{\rm WD} = 0.17\;M_{\odot}$, $P_{\rm orb} = 0.30\;$days. The blue lines represent the pre-AIC phase. 
    The solid green and dashed red lines are for the post-AIC phase with a natal kick velocity of $\omega = 0$ and $\omega = 50\;{\rm km\,s^{-1}}$, respectively. The dots indicate the immediate post-AIC systems. The arrows indicate the direction of evolution in each panel.}
    \label{fig:aic_ex1}
\end{figure*}

\begin{figure*}
    \centering
    \includegraphics[width=\columnwidth]{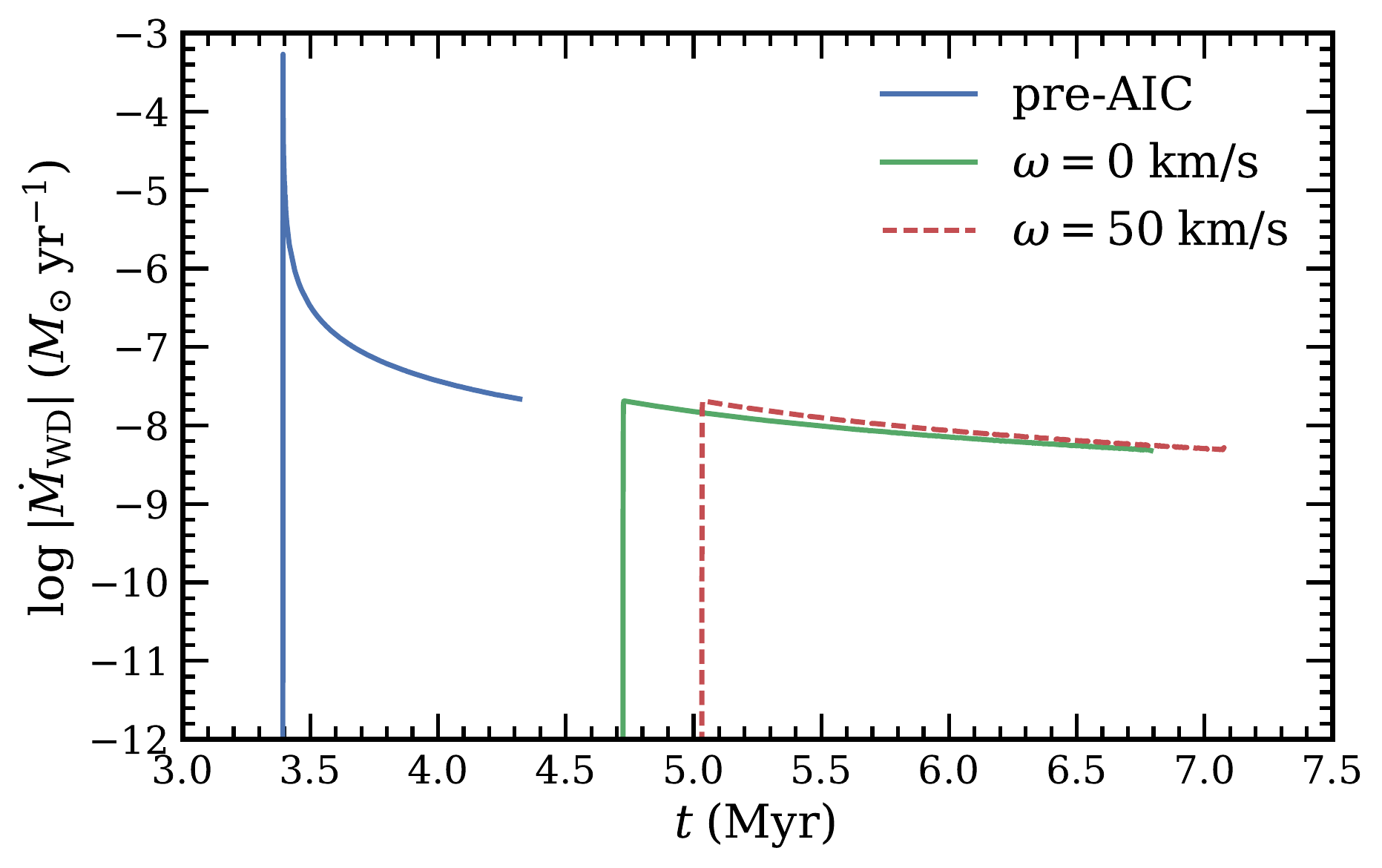}
    \includegraphics[width=\columnwidth]{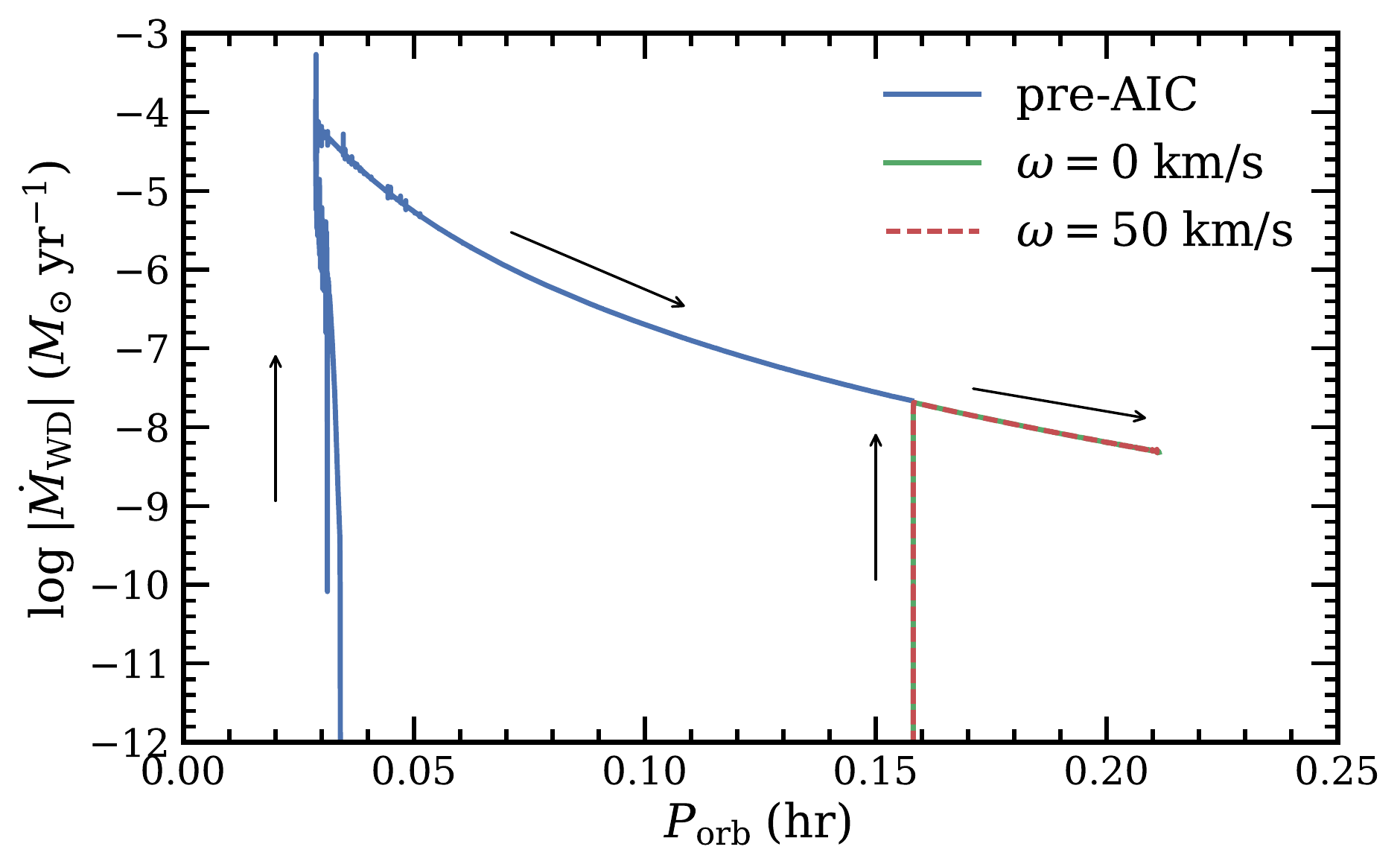}
    \includegraphics[width=\columnwidth]{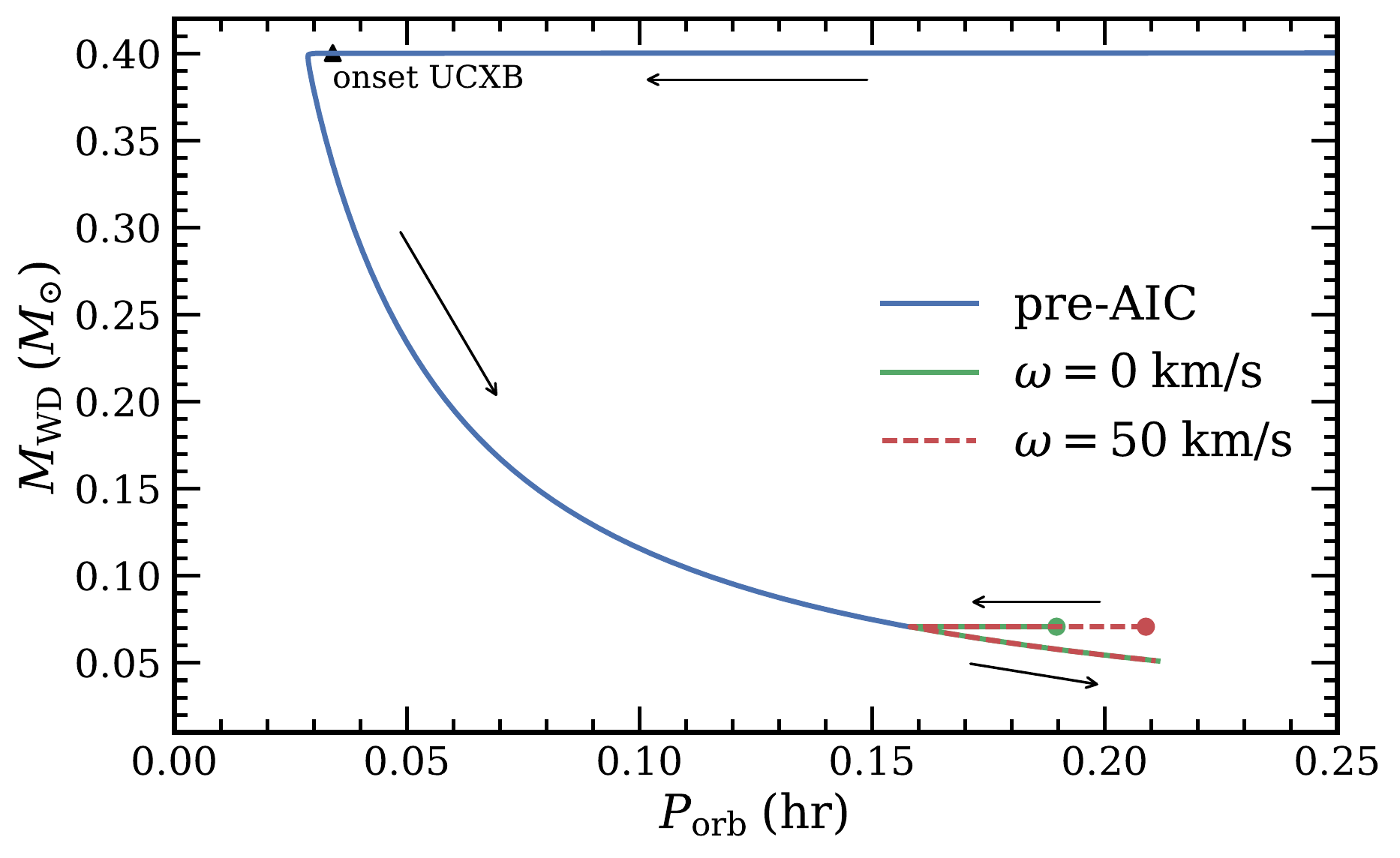}
    \includegraphics[width=\columnwidth]{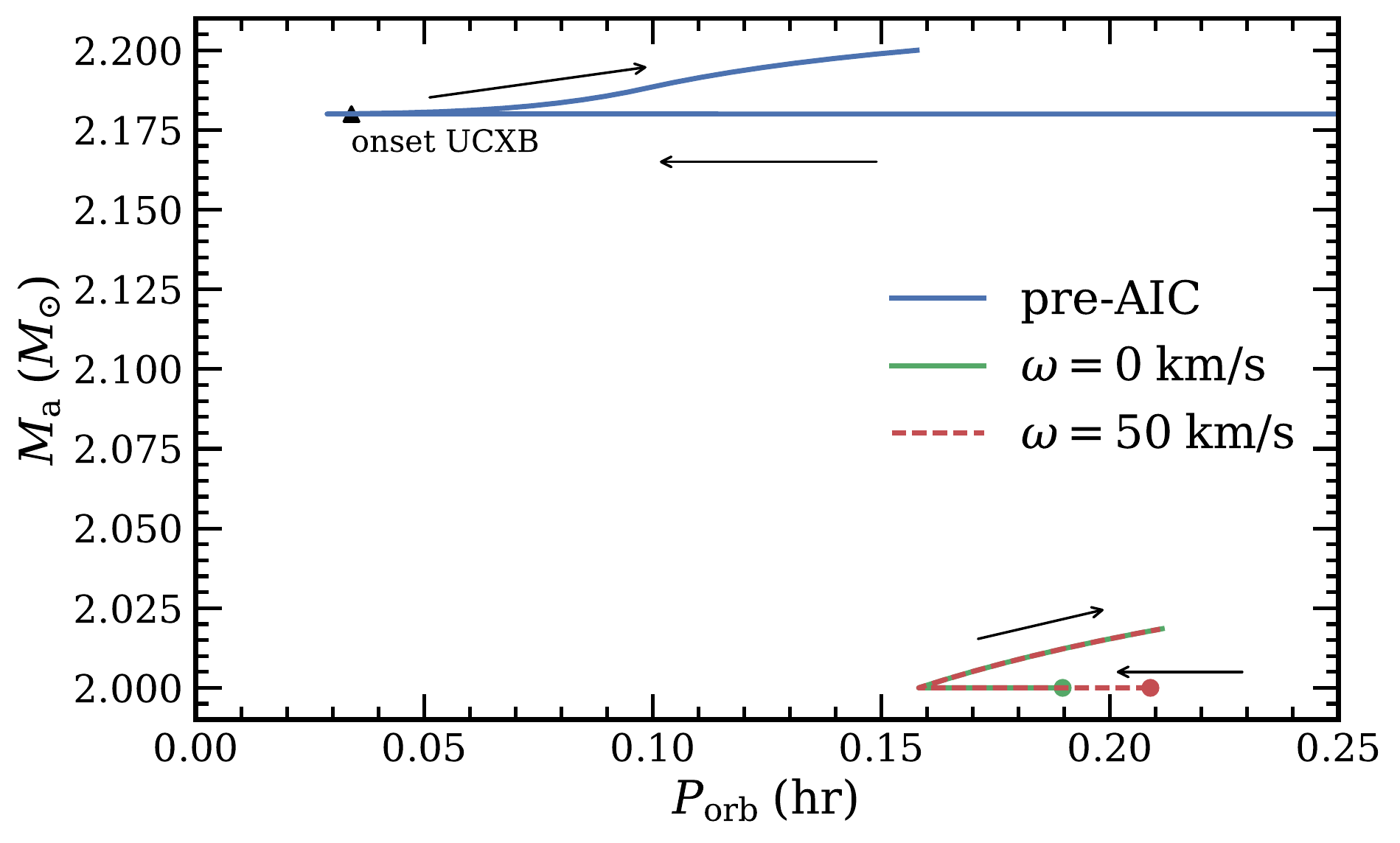}
    \caption{Similar to Fig.~\ref{fig:aic_ex1}, but here with initial binary parameters of $M_{\rm a} = 2.18\;M_{\odot}$, $M_{\rm WD} = 0.40\;M_{\odot}$ and $P_{\rm orb} = 0.02\;$days. In the upper-right panel, the mass-transfer rate is not smooth in the early phase due to numerical noise.}
    \label{fig:aic_ex2}
\end{figure*}

In Fig.~\ref{fig:aic_ex1}, we present an example of UCXB evolution leading to NS-AIC and formation of a BH-UCXB. In this example, the initial NS and He~WD masses are $2.18\;M_{\odot}$ and $0.17\;M_{\odot}$, respectively. The initial orbital period is $0.30\;$days. There are three stages in this figure: two UCXB phases (pre-AIC and post-AIC), and the intervening AIC event. In the early pre-AIC phase (blue line), prior to the UCXB stage with mass transfer, the orbital period decreases due to GW radiation. Plotted here is the last part for which $P_{\rm orb}<1.2\;{\rm hr}$. When $P_{\rm orb}$ decreases to around 0.127~hr (6.7~min), the He~WD fills its Roche lobe and starts to transfer material to the NS (the first UCXB phase), leading to an increase of the NS mass. By the time the thin H-envelope is stripped off the WD donor, the mass-transfer rate (upper panel) reaches its maximum value and the orbital period is close to its minimum value ($\sim 4.8\;{\rm min}$). After that, the orbital period starts to increase, leading to a decrease of the mass-transfer rate. When $P_{\rm orb}\simeq 0.16\;$hr and $M_{\rm WD}~\sim 0.07\;M_\odot$, the NS mass has grown to $2.20\;M_{\odot}$ and the NS is assumed to undergo AIC and produce a BH with a birth mass of $2.0\;M_{\odot}$. 

As a result of the sudden mass loss due to the NS-AIC event, the orbit widens instantaneously and the system is detached from the UCXB stage. In addition, the system may widen farther (or less, depending on kick direction) in case a kick, $w$ is applied to the newborn BH. In Fig.~\ref{fig:aic_ex1}, the red dashed line shows the post-AIC evolution for $w=50\;{\rm km\,s}^{-1}$, superimposed on top of the green line for the case of a purely symmetric AIC with $w=0$. 
In the latter case for $w=0$, the post-AIC semi-major axis and eccentricity are $0.218\;R_{\odot}$ and $0.096$, respectively. After re-circularization \citep{sut74}, the orbital period is $0.194\;$hr (see inlet in the lower left panel). 
If $w=50\;{\rm km\,s}^{-1}$, the binary semi-major axis and eccentricity after AIC are $0.238\;R_{\odot}$ and $0.172$, respectively, and the binary orbital period is $0.215\;$hr after the orbit has re-circularized.  
In both cases, the post-AIC/re-circularization separation is larger than the binary separation before the AIC event. That is the reason why the WD cannot fill its Roche lobe at first and a detached BH+WD binary is produced immediately after the AIC event. 
As the binary continues to evolve, however, the binary separation decreases due to further GW radiation. The WD fills its Roche lobe again after 0.42~Myr (0.74~Myr) in the model with $\omega = 0$ ($50\;{\rm km\,s}^{-1})$ and the binary continues to evolve as a BH-UCXB. In the second mass-transfer phase (see red dashed/green solid line in the upper panel), the mass-transfer rate is always below the Eddington limit. 

In Fig.~\ref{fig:aic_ex2}, we show another example of binary evolution with an initial WD mass of $0.40\;M_{\odot}$, and using the same initial NS mass of $2.18\;M_{\odot}$ as in the previous example. The initial orbital period is set to $0.02\;$days. The calculation is continued through the three stages of evolution until the WD is stripped down to mass of $\sim 0.05\;M_{\odot}$, at which point numerical issues lead to very small timesteps and the calculation is terminated. 
The general evolution of this binary system is very similar to the previous example shown in Fig.~\ref{fig:aic_ex1}. The main difference is that because of the much higher mass-transfer rate from this more massive WD, the evolution is faster and more material is lost from the system. Therefore, in both cases, the NS-AIC event does not occur until $M_{\rm WD}\sim 0.07\,M_\odot$ and we conclude that in NS-UCXBs with massive initial WD donors, the WD needs to lose significantly more mass in order to increase the mass of the NS to its critical value for AIC. 
This also means that NS-UCXBs with smaller initial WD masses are more likely to evolve into BH-UCXBs because systems with smaller WD masses are more abundant. 

From the upper two panels of Figs.~\ref{fig:aic_ex1} and \ref{fig:aic_ex2}, we notice that the evolution of mass-transfer rate, orbital period and WD donor mass converge to the same track for both the pre-AIC and the post-AIC phases. This is because the evolution of UCXBs depends almost exclusively on the WD donor mass, and only slightly depends on the accretor mass --- see also discussions in \citet{stli17,cthc21,ctch22b}.

\section{Discussion}\label{sec:discussion}
\subsection{Number of BH-UCXBs produced via NS-AIC}\label{subsec:number_BH-UCXBs}
Assuming that some NSs may have pre-UCXB masses very close to the maximum mass of a NS (here adopted to be $M_{\rm NS,max}=2.20\;M_{\odot}$), we have now demonstrated that UCXBs with BH accretors can be produced by AIC of a NS. In the following we will use simple back-of-the-envelope estimates to derive a rough value for the number of such BH-UCXBs produced via AIC.

In \citet{ctch22b}, we found that NS accretors with an initial mass of $1.30\;M_{\odot}$ may accrete an amount of material of up to $\Delta M_{\rm acc} \sim 0.035\;M_{\odot}$ in an UCXB with a He~WD donor of $0.17\;M_\odot$ (similar to the WD donor mass studied here). For testing the most extreme case, we repeated those calculations with a NS-UCXB with an initial NS mass of $2.18\;M_{\odot}$ and a WD mass of $0.16\;M_\odot$, and the result remained almost invariant: we found that the NS can at most accrete $0.0388\;M_{\odot}$ (if AIC is avoided). 
Therefore, for a NS-UCXB system evolving to an AIC-event, we must require that the initial NS mass in NS+He~WD binaries should be $\ge M_{\rm NS,max}-0.0388\;M_{\odot}$ (i.e. here $M_{\rm NS}\ga 2.161\;M_{\odot}$).
If we consider the possibility that NSs in UCXBs can support super-Eddington accretion, the required initial NS mass will be smaller, and thus the initial NS mass range becomes wider.  

So far, there are around 20~UCXBs known in our Galaxy. Assuming that these systems evolved from NS+He~WD systems with a flat probability density function of initial NS masses between $1.20 - 2.20\;M_{\odot}$, we expect of the order of only 1~system with an initial NS mass between $2.16-2.20\;M_{\odot}$, and therefore of the order of 1~BH-UCXB with a He~WD donor in the current sample of UCXBs. (Note that, from the upper-left panels in Figs.~\ref{fig:aic_ex1} and \ref{fig:aic_ex2}, the observable lifetime of the post-AIC BH-UCXB is much longer than that of the pre-AIC NS-UCXB, so these systems will most likely be observed with a BH.) The above estimate represents a conservative lower limit because it assumes that UCXB evolution is restricted to Eddington-limited accretion onto the NS.

\subsection{Distinguishing BH-UCXBs from NS-UCXBs}
As important GW sources, UCXBs are expected to be detectable by LISA, TianQin and Taiji. In \citet{ctch22b} (see fig.~5), we showed that the WD mass can be determined simply if the GW frequency is determined from observations. The accretor mass can then be derived from the measurement of the chirp mass (and the distance to the source can be derived from the measured GW strain). \citet{taur18} estimated that $50-100$ NS+He~WD UCXBs are expected in the LISA band. Thus we expect, as a first estimate, of the order of $2-5$ BH-UCXBs to be detected by LISA, if these systems form exclusively via NS-AIC with a low-mass He~WD donor.
However, the accretor type may be impossible to identify from GW observations alone, since the BH mass is smaller than the maximum mass of NSs in the scenario investigated here (see lower-right panels in Figs.~\ref{fig:aic_ex1} and \ref{fig:aic_ex2}). We can only point to a NS accretor with confidence if we detect electromagnetic pulsations (i.e. a radio pulsar) or a burst of X-rays (i.e. explosive thermonuclear burning from a NS surface). 
It is also possible that before the collapse of the NS in the AIC event, the NS is rapidly rotating and deformed, leading to an emission of a significant continuous high-frequency GW signal \citep{and19}. However, as mentioned previously, the pre-AIC NS-UCXB stage is short-lived, and so far, no continuous GW signal has been detected from any sources by LIGO-Virgo-KAGRA \citep{aaa-LIGO+22}.
For UCXBs with accretor masses exceeding $\sim 3\;M_\odot$, the accretor must be a BH, and it must be produced via another scenario than investigated here.

\subsection{Signatures from the NS-AIC event}
\citet{da06} suggested that sufficient energy can be released through $\nu\;\bar{\nu}\rightarrow{e}^{+} e^{-}$ processes to power a short gamma-ray burst during the AIC of a NS. Given the long delay time of NS+He~WD binaries (and also the long timescale expected from the subsequent GW shrinking of the orbit to produce the UCXB models studied here), this kind of AIC-driven short gamma-ray burst would therefore be likely to appear in an old stellar population. 
This is in accordance with the finding that the progenitors of short gamma-ray bursts have a preference for early-type galaxies \citep{gnoc+08}.
In addition, \citet{hy13} found that the typical delay time of short gamma-ray bursts is $\tau > 3\;{\rm Gyr}$. 

\citet{fr14} suggested that the surface of the NS will be hidden behind the event horizon of the BH during the collapse and the NS magnetosphere will be disrupted violently. This will produce strong emission in the radio regime, possibly observable as a non-repeating fast radio burst \citep[see, however,][for a recent review of other models for fast radio bursts]{phl22}.

\subsection{X-ray luminosity}\label{subsec:Xrays}
\begin{figure}
    \centering
    \includegraphics[width=\columnwidth]{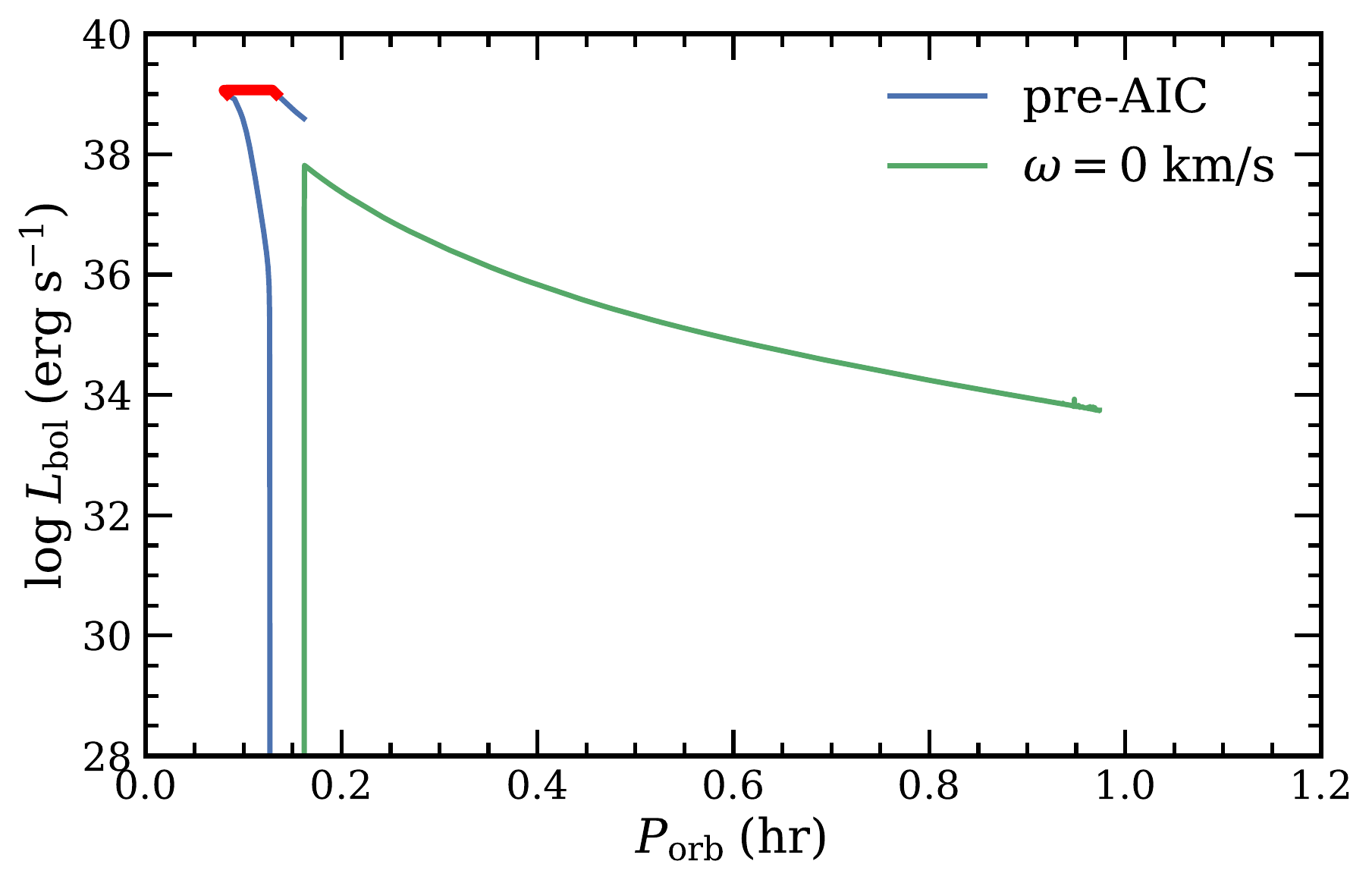}
    \caption{Evolution of bolometric X-ray luminosity as a function of orbital period for the binary system presented in Fig.~\ref{fig:aic_ex1}. Here we only present the results for the case with $\omega = 0$, since the results for $\omega = 50\;{\rm km\,s}^{-1}$ are similar. 
    Shortly after the onset of the UCXB phase, and prior to the AIC, the binary would be observed as an ultraluminous X-ray source, i.e. $L_{\rm bol} > 10^{39}\;{\rm erg\,s}^{-1}$ (indicated with red color; see also Section~\ref{subsec:Xrays}).}
    \label{fig:lum_evl}
\end{figure}

Following \citet{mkfl+22}, see their eqs.(1-4), one can estimate the bolometric X-ray luminosity of UCXBs. In Fig.~\ref{fig:lum_evl}, we present the evolution of bolometric X-ray luminosity for the binary system presented in Fig.~\ref{fig:aic_ex1}. From this plot, we find that the bolometric X-ray luminosity of the long-orbital period post-AIC UCXBs generally is larger than $10^{34}\;{\rm erg\,s}^{-1}$.
At short orbital periods, the X-ray bolometric luminosity is significantly larger. For example, around the minimum orbital period, the bolometric X-ray luminosity of the pre-AIC system even exceeds $\sim 10^{39}\;{\rm erg\,s}^{-1}$ and the system may appear as an ultraluminous X-ray source. The duration of this phase is around $4.4 \times 10^{5}\;{\rm yr}$, i.e. comparable to some of the intermediate-mass X-ray binary candidates as ultraluminous X-ray sources studied by \citet{mft+20}. Given their long delay time, these NS-UCXBs (and also other NS-UCXBs that do not undergo AIC) may explain some of the ultraluminous X-ray sources observed in old stellar population environments \citep[e.g.][]{iba04}. 

\subsection{Formation of BHs in the mass gap}
It is commonly suggested that there is a mass gap between the lowest mass BHs ($\sim5.0\;M_{\odot}$) and the maximum mass of NSs ($\sim 2-3\;M_{\odot}$, \citealt{bjco98,opnm10,fsck+11}). In our simulations, the birth mass of the BH following an AIC event is below the maximum mass of a NS because of the assumed release of gravitational binding energy of $0.20\;M_\odot\,c^2$. Its mass grows during the BH-UCXB phase but it remains below the maximum mass of NSs simply because the mass-transfer rate is small (Figs.~\ref{fig:aic_ex1} and \ref{fig:aic_ex2}).
Therefore, we do not expect that the NS-AIC scenario will produce X-ray binaries with a BH mass in the mass gap {\em if} the WD donor mass is low enough $\lesssim 0.45\;M_\odot$ to secure a stable RLO \citep{ctch22b}. 
On the other hand, the NS+WD binaries with more massive WDs are likely to undergo dynamically unstable mass transfer and eventually merge during their UCXB phase. The NSs in these systems may indeed collapse into BHs with masses in the mass gap. However, such BHs will be isolated and thus difficult to detect (let alone a mass measurement).

\subsection{Influence of the NS spin}
During the mass-transfer phase, the NS can be spun up as it accretes material from the WD donor. It is known that spin will influence the maximum mass of a NS.
The maximum mass of a uniformly rotating NS is $\sim 1.20\pm{0.02}$ times the maximum mass of a non-rotating NS \citep{br16}.
In our calculations, the NS only accretes $\sim 0.02\;M_{\odot}$ before it collapses to a BH. The equilibrium spin period of a NS accreting this amount can be estimated with the following expression \citep{tlk12}:
\begin{equation}
  \Delta M_{\rm acc} = 0.22\;M_\odot \; \frac{(M_{\rm NS}/M_\odot)^{1/3}}{P_{\rm ms}^{4/3}},
\end{equation}  
where $\Delta M_{\rm acc}$ is the mass accreted by the NS, and $P_{\rm ms}$ is the equilibrium spin period of the NS in units of milliseconds. 
Thus we expect a spin period of $\sim 7.4\;{\rm ms}$, which is much longer than the break-up limit of $\sim 0.6\;{\rm ms}$. As a result, the effect of NS spin on the maximum mass of the NS is rather small in our calculations \citep{br16,mpwr20}, and any delay of the AIC due to rapid NS spin can be disregarded. If we adopt a NS spin period of $P_{\rm ms}\sim 7.4$ and assume no angular momentum loss during the AIC, we get an upper limit of the spin parameter of the resultant BH, $a_\ast = Jc/(GM^2) \sim 0.05$. This means that the resultant BH has a very small spin.

\subsection{Comparison with other's work}
\label{subsec:com}
Recently \citet{skkb+22} modelled the evolution of a population of LMXBs with H-rich main-sequence donors. They found that some NSs may collapse into BHs in such binaries, thereby producing BH-LMXBs with BH masses of $3-5\;M_{\odot}$  (see their fig.~2). A main difference from our work (besides the nature of the X-ray binaries investigated: LMXBs vs UCXBs) is that in their simulations they assumed the mass transfer to be either conservative or applying an accretion efficiency of 50\% whenever the mass-transfer rate is sub-Eddington. We have applied an accretion efficiency of 30\% (a conservative upper limit) based on the firm arguments provided in the references given in Section~\ref{subsec:aim}. 
Furthermore, \citet{skkb+22} allowed for a maximum NS birth mass of $3.0\;M_{\odot}$ and donor stars up to $3.0\;M_
\odot$. The combination of these three effects is the reason for the possibility to produce BHs between $3-5\;M_\odot$. 
The BH-LMXBs produced via AIC of NSs in their simulations mainly have orbital periods larger than $1.0\;$hr. It is unclear whether they continued their evolution to include UCXBs in their population. 
In addition, they claimed that the maximum birth mass of NSs has to be $< 2.0\;M_{\odot}$ in order to ``suppress the formation of AIC LMXBs''. Based on the provided information and methodology, we are unable to judge whether this is a valid conclusion, but we notice that the aforementioned accretion efficiency {\it is} important for LMXBs and that additional effects, like possible mass loss during the AIC event, may well combine to affect their conclusion.

\citet{gls22} modelled the evolution of LMXBs and intermediate X-ray binaries (IMXBs) assuming the possibility of highly super-Eddington accretion that obviously plays a critical role in mass growth of NSs. They also found that BH X-ray binaries can be produced via the AIC of NSs in L/IMXBs --- even for $M_{\rm NS,max}=2.5-3.0\;M_\odot$ (thus creating a formation channel for lower mass-gap BHs). In their work, however, they did not consider the evolution of UCXBs with WD donors. 
In our calculations, we did not include the possibility of super-Eddington accretion given the doubt of its existence that depends on the interpretation of the ultra-luminous X-ray sources and the question of beaming \citep[e.g.][]{kfr17}{\footnote{ We notice that there is some indirect evidence for super-Eddington accretion in massive X-ray binaries, e.g. \citet{vdbr+20} and \citet{bse23}.}. For instance, beaming of the emitted radiation would imply that the actual full-sky luminosity of these sources is much lower than their apparent isotropic luminosity. 

\subsection{Final remarks on the uncertainty of the maximum mass of a NS}
Based on the summary and references in Section~\ref{subsec:aim}, we believe that a maximum NS mass of $2.20\;M_{\odot}$ is most realistic. However, we can not exclude the possibility of a larger value (e.g. $2.60\;M_\odot$). As long as the NS can have a birth mass very close to the maximum mass of NS, the scenario we propose in this paper should still work. 
Recent 3D modelling of SNe by e.g. \citet{brv+20,pm20} support that NSs should have a continuous birth-mass distribution, indicating the possibility of NSs with a birth mass very close to the maximum mass \citep[see also the expectations from the {\it delayed explosion prescription} of][]{fbwd+12}.

\section{Conclusions}\label{sec:conclusion}
Given that some NSs in binary MSP systems may have masses close to the maximum possible mass of a NS, we have investigated the possibility of formation of BH-UCXBs via AIC of an accreting NS in a NS-UCXB. We calculated detailed evolutionary tracks of a few selected UCXBs before and after the AIC event to probe the final properties of such BH-UCXBs. We evolved UCXBs with He~WD donors of $0.17\;M_\odot$ and $0.40\;M_\odot$. The former case represents NS+He~WD UCXBs produced via stable RLO in LMXBs, which always produce He-WD donors of $0.15-0.17\;M_\odot$ \citep{taur18,cthc21}, and the latter case is an example with a relatively massive WD donor (e.g. representing an example of an UCXB produced via CE evolution). We modelled these AIC events with and without a natal NS kick. The main results are summarized as follows:

\begin{enumerate}

\item We find that only NS-UCXBs originating from LMXBs with a NS mass close to the maximum mass of a NS (here assumed to be $2.20\;M_\odot$) may evolve into BH-UCXBs after the NS accretes sufficient material from its WD companion to trigger an AIC. Based on the evidence of rather inefficient accretion onto a NS (as supported from observations of recycled pulsar masses), 
we conclude that a NS with any mass can at most accrete $\Delta M_{\rm acc}=0.039\;M_\odot$ during the UCXB phase \citep[in agreement with][]{ctch22b}.
This small amount of accreted material limits the AIC scenario to UCXBs with a NS born within $\Delta M_{\rm acc}$ of the maximum NS mass limit (e.g. NSs born with masses $>2.161\;M_\odot$).

\item The modest accretion onto the NS/BH in an UCXB has two consequences: (i) it means that the NS spin-up is limited (final spin periods of above $\sim 7\;{\rm ms}$) and therefore that rotation will not affect the NS equation-of-state and its maximum mass at any significant level; and (ii) the post-AIC BH will have a mass throughout the subsequent BH-UCXB phase that remains below that of the NS at the moment of the AIC (because of release of gravitational binding energy during the AIC implosion).

\item The evolution of post-AIC BH-UCXBs is very similar to the evolution of NS-UCXBs. The only major difference is that the AIC event causes an instantaneous widening of the orbit that results in a temporary detachment of the WD donor star from its Roche lobe. Thus following the AIC, the system is unobservable in X-rays for $0.4-0.7\;{\rm Myr}$ --- the time interval depends on the amount of orbital widening that might include the effect from a potential NS birth kick.

\item We expect that there is of the order of one BH-UCXB with a He~WD companion in current known sample of UCXBs and that of the order $2-5$ BH-UCXBs can be detected by LISA. A firm identification of the accretor being a BH is non-trivial, whereas an EM or GW signature may prove the presence of a NS accretor.

\item The X-ray luminosity of NS-UCXBs near their orbital period minimum, shortly after the onset of RLO, exceeds $\sim 10^{39}\;{\rm erg\,s}^{-1}$ and the system may thus appear as an ultraluminous X-ray source. The duration of this phase from the test model in Fig.~\ref{fig:aic_ex1} is around $4.4 \times 10^{5}\;{\rm yr}$. The ability of NS-UCXBs to be observable as ultraluminous X-ray sources is independent of whether they subsequently undergo AIC or not.

\end{enumerate}

\begin{acknowledgments}
We thank the anonymous referees for comments that helped improving the manuscript.
We thank John Antoniadis and Savvas Chanlaridis for discussions on the influence of NS spin, and Ed van den Heuvel for discussions on AIC. This work is partially supported by the National Key R\&D Program of China (grant Nos. 2021YFA1600403, 2021YFA1600401), the National Natural Science Foundation of China (Grant No. 12090040/12090043, 12073071, 11873016, 11733008, 12125303),
Yunnan Fundamental Research Projects (Grant No. 202001AT070058, 202101AW070003,202201BC070003), the “Yunnan Revitalization Talent Support Program"—Science \& Technology Champion Project (NO. 202305AB350003), the science research grants from the China Manned Space Project with No. CMS-CSST-2021-A10 and the CAS light of West China Program. The authors gratefully acknowledge the ''PHOENIX Supercomputing Platform'' jointly operated by the Binary Population Synthesis Group and the Stellar Astrophysics Group at Yunnan Observatories, CAS.
We are grateful to the \textsc{mesa} council for the \textsc{mesa} instrument papers and website. 

\end{acknowledgments}

\vspace{5mm}
\software{MESA \citep{pbdh+11,pcab+13,pmsb+15,psbb+18,pssg+19}, matplotlib \citep{hunt07}, numpy \citep{numpy20}, astropy \citep{astropy13,astropy18},
Python from \dataset[python.org]{https://www.python.org/}}. 

\bibliography{sample631,hailiang_refs}{}
\bibliographystyle{aasjournal}

\end{document}